AIAA-2007-6201

# Optimal Electrostatic Space Tower (Mast, New Space Elevator)*

**Alexander A. Bolonkin**
C & R, 1310 Avenue R, Suite 6-F
Brooklyn, New York 11229, USA
aBolonkin@juno.com, http://Bolonkin.narod.ru

**Abstract**

Author offers and researched the new and revolutionary inflatable electrostatic AB space towers (mast, new space elevator) up to one hundred twenty thousands kilometers (or more) in height.

The main innovation is filling the tower by electron gas, which can create pressure up one atmosphere, has negligible small weight and surprising properties.

The suggested mast has following advantages in comparison with conventional space elevator: 1. Electrostatic AB tower may be built from Earth's surface without the employment of any rockets. That decreases the cost of electrostatic mast by thousands of times. 2. One can have any height and has a big control load capacity. 3. Electrostatic tower can have the height of a geosynchronous orbit (36,000 km) WITHOUT the additional top cable as the space elevator (up 120,000 ÷ 160,000 km) and counterweight (equalizer) of hundreds of tons. 4. The offered mast has less total mass than conventional space elevator. 5. The offered mast can be built from less strong material than space elevator cable. 6. The offered tower can have the high-speed electrostatic climbers moved by high-voltage electricity from Earth's surface. 7. The offered tower is safer resisting meteorite strikes than an ordinary cable space elevator. 8. The electrostatic mast can bend in any needed direction when we give the necessary electric voltage in the required parts of the extended mast. 9. Control mast has stability for any altitude. Three projects 100 km, 36,000km (GEO), 120,000 km are computed and presented.

These towers can be used for tourism, scientific observation of space, observation of the Earth's surface, weather and upper atmosphere experiments, and for radio, television, and communication transmissions. These towers can also be used to launch interplanetary spaceships and Earth-orbiting satellites.

**Key words**: Space tower, electrostatic space mast, space tourism, space communication, space launch, space observation

## Introduction

**Brief History.** The idea of building a tower high above the Earth into the heavens is very old [1]. The writings of Moses, in chapter 11 of his book *Genesis* refers to an early civilization that tried to build a tower to heaven out of brick and tar. This construction was called the Tower of Babel, and was reported to be located in Babylon in ancient Mesopotamia. Later in chapter 28, Jacob had a dream about a staircase or ladder built to heaven. This construction was called Jacob's Ladder. More contemporary writings on the subject date back to K.E. Tsiolkovski in his manuscript "Speculation about Earth and Sky and on Vesta," published in 1895 [2]. This idea inspired Sir Arthur Clarke to write his novel, *The Fountains of Paradise* [3], about a space tower (elevator) located on a fictionalized Sri Lanka, which brought the concept to the attention of the entire 20th Century world.

Today, the world's tallest construction is a television transmitting tower (mast) near Fargo, North Dakota, USA. It stands 629 m high and was built in 1963 for KTHI-TV. The CNN Tower in Toronto, Ontario, Canada is the world's tallest building. It is 553 m in height, was completed in1975, and has the world's highest observation deck at 447 m. The tower structure is concrete up to the observation deck level. Above is a steel structure supporting radio, television, and communication antennas. The total weight of the tower is 3,000,000 metric tons.

The Ostankin Tower in Moscow is 540 m in height and has an observation desk at 370 m. The world's tallest office building is the Petronas Towers in Kuala Lumpur, Malasia. The twin towers are 452 m in height. They are 10 m taller than the Sears Tower in Chicago, Illinois, USA. The Skyscrapers (Taipei, Taiwan, 2004) has height of 509 m, the Eiffel Tower (Paris, 1887-1889) has 300 m, Empire State Building (USA, New York, 1930-1931) has 381 m + TV mast of 61 m. Under construction a building of 1001 m (Kuwait City, Kuwait) and 1430 m Supported Structure in Gulf of Mexico.

Current materials make it possible even today to construct towers many kilometers in height. However, conventional towers are very expensive, costing billions of dollars. When considering how high a tower can be built, it is important to remember that it can be built to high height if the base is large enough. Theoretically, you could build a tower to geosynchronous Earth orbit (GEO) out of bubble gum, but the base would likely cover half the surface of the Earth.

**The new types of towers**. The author offered and researched a series on new towers (masts) [6]-[11]: optimal inflatable towers filled by gas (air, helium, hydrogen), optimal solid towers, new kinetic cable towers.

The offering new revolutionary electrostatic tower is based on old (1982) ideas author of using electrostatic forces [4]-[5]. They are applied to space tower and are shown the gigantic advantages in comparison with conventional space elevator. Some of these advantages named in abstract over. Main of them are follow: electrostatic mast can be built any height without rockets, one needs material in tens times less them space elevator. That means the electrostatic mast will be in hundreds times cheaper then conventional space elevator. One can be built on the Earth's surface and their height can be increased as necessary. Their base is very small.

The main innovations in this project are the application of electron gas for filling tube at high altitude and a solution of a stability problem for tall (thin) inflatable mast by control structure.

**The tower applications**. The high towers (3-100 km) have numerous applications for government and commercial purposes:

- Entertainment and Observation platform.
- Entertainment and Observation desk for tourists. Tourists could see over a huge area, including the darkness of space and the curvature of the Earth's horizon.
- Drop tower: tourists could experience several minutes of free-fall time. The drop tower could provide a facility for experiments.
- A permanent observatory on a tall tower would be competitive with airborne and orbital platforms for Earth and space observations.
- Communication boost: A tower tens of kilometers in height near metropolitan areas could provide much higher signal strength than orbital satellites.
- Solar power receivers: Receivers located on tall towers for future space solar power systems would permit use of higher frequency, wireless, power transmission systems (e.g. lasers).
- Low Earth Orbit (LEO) communication satellite replacement: Approximately six to ten 100-km-tall towers could provide the coverage of a LEO satellite constellation with higher power, permanence, and easy upgrade capabilities.

The towers having a height 36,000 ÷ 120,000 km may be used for free launching the Earth's satellites and interplanetary ships and as space station for arriving space ships.
Other new revolutionary methods of access to space are described in [10]-[14].

## Description of Installation and Innovations

**1. Electrostatic tower**. The offered electrostatic space tower (or mast, or space elevator) is shown in fig.1. That is inflatable cylinder (tube) from strong thin dielectric film having variable radius. The film has inside the sectional thin conductive layer 9. Each section is connected with issue of control electric voltage. In inside the tube there is the electron gas from free electrons. The electron gas is separated by in sections by a thin partition 11. The layer 9 has a positive charge equals a summary negative charge of the inside electrons. The tube (mast) can have the length (height) up Geosynchronous Earth Orbit (GEO, about 36,000 km) or up 120,000 km (and more) as in our project (see computation below). The very high tower allows to launch free (without spend energy in launch stage) the interplanetary space ships. The offered optimal tower is design so that the electron gas in any cross-section area compensates the tube weight and tube does not have compressing longitudinal force from weight. More over the tower has tensile longitudinal (lift) force which allows the tower has a vertical position. When the tower has height more GEO the additional centrifugal force of the rotate Earth provided the vertical position and natural stability of tower.

The bottom part of tower located in troposphere has the bracing wires 4 which help the tower to resist the troposphere wind.

The control sectional conductivity layer allows to create the high voltage running wave which accelerates (and brakes) the cabins (as rotor of linear electrostatic engine [11]) to any high speed. Electrostatic forces also do not allow the cabin to leave the tube.

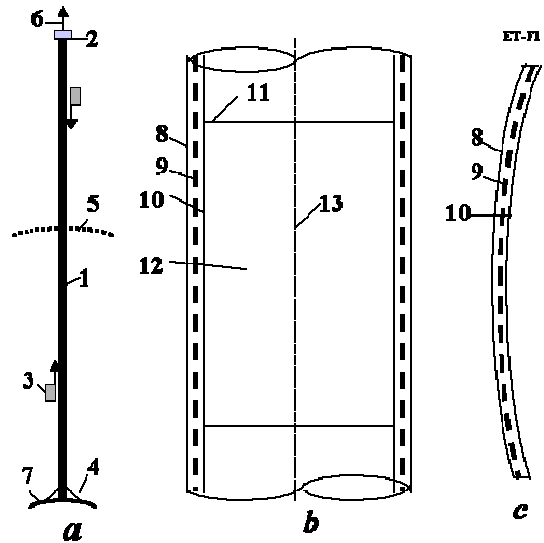

**Fig.1**. Electrostatic AB tower (mast, Space Elevator). (a) Side view, (b) Cross-section along axis, (c) Cross-section wall perpendicular axis. Notation: 1 - electrostatic AB tower (mast, Space Elevator); 2 - Top space station; 3 - passenger, load cabin with electrostatic linear engine; 4 - bracing (in troposphere); 5 - geosynchronous orbit; 6 - tensile force from electron gas; 7 - Earth; 8 - external layer of isolator; 9 - conducting control layer having sections; 10 - internal layer of isolator; 11 - internal dielectric partition; 12 - electron gas, 13 - laser control beam.

**2. Electron gas and AB tube**. The electron gas consists of conventional electrons. In contract to molecular gas the electron gas has many surprising properties. For example, electron gas (having same mass density) can have the different pressure in the given volume. Its pressure depends from electric intensity, but electric intensity is different in different part of given volume (fig.2b). For example, in our tube the electron intensity is zero in center of cylindrical tube and maximum at near tube surface.

The offered AB-tube is main innovation in the suggested tower. One has a positive control charges isolated thin film cover and electron gas inside. The positive cylinder create the zero electric field inside the tube and electron conduct oneself as conventional molecules that is equal mass density in any points. When kinetic energy of electron is less then energy of negative ionization of the dielectric cover or the material of the electric cover does not accept the negative ionization, the electrons are reflected from cover. In other case the internal cover layer is saturated by negative ions and begin also to reflect electrons. Impotent also that the offered AB electrostatic tube has neutral summary charge in outer space.

**Advantages of electrostatic tower**. The offered electrostatic tower has very important advantages in comparison with space elevator:
1. Electrostatic AB tower (mast) may be built from Earth's surface without rockets. That decreases the cost of electrostatic mast in thousands times.
2. One can have any height and has a big control load capacity.

3. In particle, electrostatic tower can have the height of a geosynchronous orbit (37,000 km) WITHOUT the additional continue the space elevator (up 120,000 ÷ 160,000 km) and counterweight (equalizer) of hundreds tons [10], Ch.1.
4. The offered mast has less the total mass in tens of times then conventional space elevator.
5. The offered mast can be built from lesser strong material then space elevator cable (comprise the computation here and in [10] Ch.1).
6. The offered tower can have the high speed electrostatic climbers moved by high voltage electricity from Earth's surface.
7. The offered tower is more safety against meteorite then cable space elevator, because the small meteorite damaged the cable is crash for space elevator, but it is only create small hole in electrostatic tower. The electron escape may be compensated by electron injection.
8. The electrostatic mast can bend in need direction when we give the electric voltage in need parts of the mast.

The electrostatic tower of height 100 ÷ 500 km may be built from current artificial fiber material in present time. The geosynchronous electrostatic tower needs in more strong material having a strong coefficient $K \geq 2$ (whiskers or nanotubes, see below).

### 3. Other applications of offered AB tube idea.

The offered AB-tube with the positive charged cover and the electron gas inside may find the many applications in other technical fields. For example:

1) *Air dirigible*. (1) The airship from the thin film filled by an electron gas has 30% more lift force then conventional dirigible filled by helium. (2) Electron dirigible is significantly cheaper then same helium dirigible because the helium is very expensive gas. (3) One does not have problem with changing the lift force because no problem to add or to delete the electrons.
2) *Long arm*. The offered electron control tube can be used as long control work arm for taking the model of planet ground, rescue operation, repairing of other space ships and so on [10] Ch.9.
3) *Superconductive or closed to superconductive tubes*. The offered AB-tube must have a very low electric resistance for any temperature because the electrons into tube to not have ions and do not loss energy for impacts with ions. The impact the electron to electron does not change the total impulse (momentum) of couple electrons and electron flow. If this idea is proved in experiment, that will be big breakthrough in many fields of technology.
4) *Superreflectivity*. If free electrons located between two thin transparency plates, that may be superreflectivity mirror for widely specter of radiation. That is necessary in many important technical field as light engine, multy-reflect propulsion [10] Ch.12 and thermonuclear power [15].

The other application of electrostatic ideas is Electrostatic solar wind propulsion [10] Ch.13, Electrostatic utilization of asteroids for space flight [10] Ch.14, Electrostatic levitation on the Earth and artificial gravity for space ships and asteroids [14, 10 Ch.15], Electrostatic solar sail [10] Ch.18, Electrostatic space radiator [10] Ch.19, Electrostatic AB ramjet space propulsion [14], etc.

## Theory and Computation

Below reader find the evidence of main equations, estimations, and computations.

**1. Optimal radius (cross-section) area of tower**. Assume we have tower from thin film filled by electron gas. Take the thin ring of tower cover with $dH$ height (Fig.2a). For getting the optimal radius the weight (force in N) $g\gamma\sigma dL$ of this elementary ring must be support by electron gas pressure $pdr$. From projection of force on vertical axis we have

$$pdr = g\gamma\delta dL, \quad dL \approx dH, \quad pdr = g\gamma\delta dH, \quad (1)$$

where $p$ is electron (charge) pressure, N/m$^2$; $dr$ and $dH$ is elementary radius and tower height respectively (see fig.2), m; $g$ is Earth gravity at altitude $H$, m/s$^2$; $\gamma$ is cover density, kg/m$^3$; $\delta$ is cover thickness, m.

The gravity for rotated Earth and electron (charge) pressure are (see [10] Ch.1)

$$g = g_0\left[\left(\frac{R_0^2}{R}\right)^2 - \frac{\omega^2 R}{g_0}\right], \quad p = \frac{\varepsilon_0 E^2}{2}. \tag{2}$$

where $g_0 = 9.81$ m/s$^2$ is Earth's gravity at altitude $H = 0$; $R_0 = 6378$ km is radius of Earth, m; $R = R_0 + H$ is distance from given cross-section tower to center of Earth, m; $\omega = 72.685 \times 10^{-6}$ rad/s is angle speed of the Earth; $E$ is maximum electric intensity, V/m (fig.2b); $\varepsilon_0 = 8.85 \times 10^{-12}$ F/m is electrostatic constant.

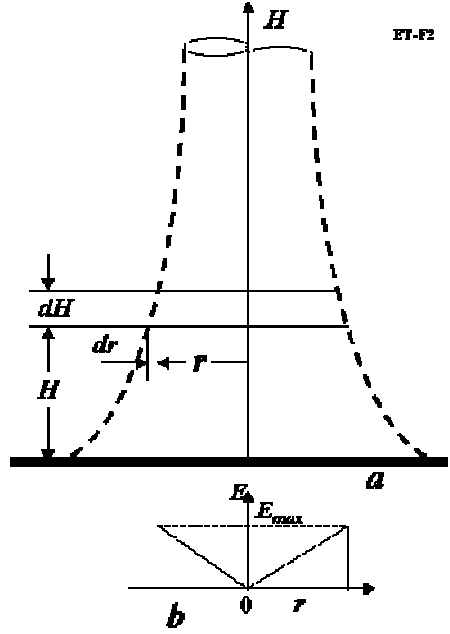

**Fig. 2**. (a) For explanation of theory optimal cross-section area of the electrostatic AB tower. (b) graph of electric intensity into tower

Look your attention that electron gas is different from conventional molecular gas. That can have a different electric intensity (that means a different pressure!) in different place of volume. The electron pressure equals zero in axis of tube and one is maximum at maximum radius of tube.

In optimal tower the electronic pressure must keep the cover

$$2rpdL = 2\delta\sigma dL \quad \text{or} \quad \delta = \frac{rp}{\sigma} \quad \text{or} \quad \delta = \frac{r\varepsilon_0 E^2}{2\sigma} \quad \text{or} \quad \bar{\delta} = \frac{\delta}{r} = \frac{\varepsilon_0 E^2}{2\sigma}, \tag{3}$$

Substitute (2)-(3) in (1) and integrate we receive

$$\int_{-r_0}^{-r}\frac{dr}{r} = \frac{g_0}{k}\int_{R_0}^{R}\left[\left(\frac{R_0^2}{R}\right)^2 - \frac{\omega^2 R}{g_0}\right]dR \quad \text{or} \quad \bar{r} = \frac{r}{r_0} = \exp\left\{-\frac{g_0 R_0^2}{k}\left[\left(\frac{1}{R_0} - \frac{1}{R}\right) - \frac{\omega^2}{2g_0}\left(\frac{R^2}{R_0^2} - 1\right)\right]\right\}, \tag{4}$$

where $k = \sigma/\gamma$ is coefficient relative strength, m/s, $K = k/10^7$.

The computation equation (4) via $H$ for different $K$ are presented in fig. 3.

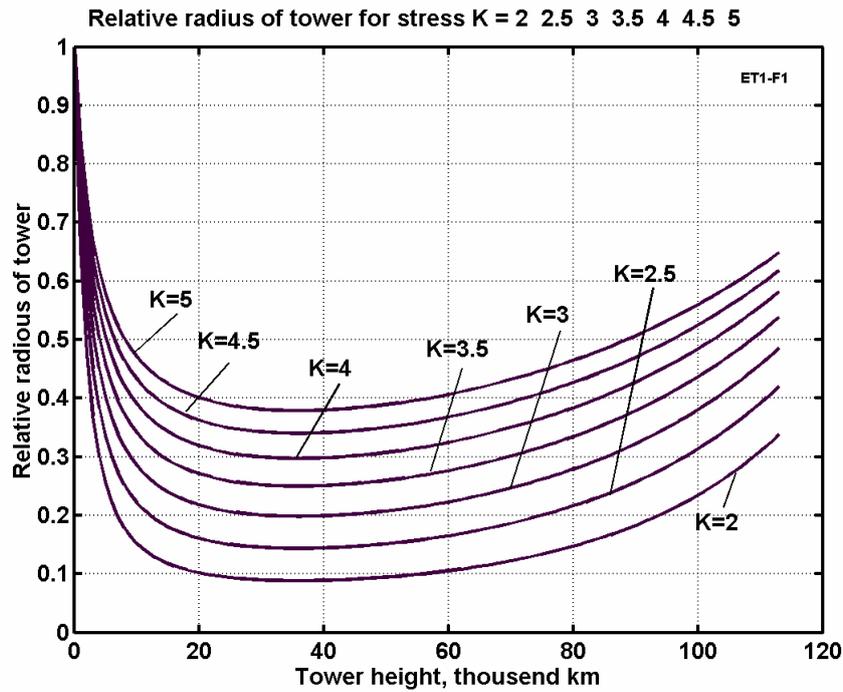

**Fig. 3.** Relative radius of electrostatic tower versus height and strong of cover film. $K = k/10^7$.

As you see than more a relative strength of cover then is more the tower diameter at geosynchronous orbit (36,000 km) and then more the lift force of tower everywhere at *H* for given *p*. In difference of space elevator the electrostatic AB tower may be built for small $K < 2$. But the ratio $S_o/S_{gco}$ in this case is big (here *S* is area of tower base and cross-section area of tower at geosynchronous orbit respectively).

**2. Material strength**. Let us consider the following experimental and industrial fibers, whiskers, and nanotubes [16]-[19]:
1. Experimental nanotubes CNT (carbon nanotubes) have a tensile strength of 200 Giga-Pascals (20,000 kg/mm$^2$). Theoretical limit of nanotubes is 30,000 kg/mm$^2$.
2. Young's modulus is over 1 Tera Pascal, specific density $\gamma = 1800$ kg/m$^3$ (1.8 g/c$^3$) (year 2000).
   For safety factor $n = 2.4$, $\sigma = 8300$ kg/mm$^2$ = $8.3 \times 10^{10}$ N/m$^2$, $\gamma = 1800$ kg/m$^3$, $k = (\sigma/\gamma) = 46 \times 10^6$, $K = 4.6$. The SWNTs nanotubes have a density of 0.8 g/cm$^3$, and MWNTs have a density of 1.8 g/cm$^3$ (average 1.34 g/cm$^3$). Unfortunately, the nanotubes are very expensive at the present time. They cost is about \$100 g (2004).
3. For whiskers $C_D$ $\sigma = 8000$ kg/mm$^2$, $\gamma = 3500$ kg/m$^3$ (1989) [16 or 10, p. 33], $n = 1$, $K_{max} = 2.37$. Cost is about \$400/kg (2001).
4. For industrial fibers $\sigma = 500 \div 600$ kg/mm$^2$, $\gamma = 1800$ kg/m$^3$, $\sigma/\gamma = 2{,}78 \times 10^6$, $n = 1$, $K_{max} = 0.28$. Cost is about $2 \div 5$ \$/kg (2003).

Figures for some other experimental whiskers and industrial fibers are given in Table 1.

Table 1. Tensile strength and density of whiskers and fibers

| Material Whiskers | Tensile strength kg/mm$^2$ | Density g/cm$^3$ | Material Fibers | Tensile strength kg/mm$^2$ | Density g/cm$^3$ |
|---|---|---|---|---|---|
| AlB$_{12}$ | 2650 | 2.6 | QC-8805 | 620 | 1.95 |
| B | 2500 | 2.3 | TM9 | 600 | 1.79 |
| B$_4$C | 2800 | 2.5 | Thoroel | 565 | 1.81 |
| TiB$_2$ | 3370 | 4.5 | Alien 1 | 580 | 1.56 |
| SiC | 2100-4140 | 3.22 | Alien 2 | 300 | 0.97 |
| Al oxide | 2800-4200 | 3.96 | Kevlar | 362 | 1.44 |

See Reference [10] p. 33.

**3. Useful lift force**. The useful (tensile) lift force of AB tower may be computed by equation

$$F = p_a S r^2, \quad p_a = \frac{1}{2} p = \frac{\varepsilon_0 E^2}{4}, \quad F = \frac{\pi \varepsilon_0 E^2}{4} r^2 \bar{r}^2, \quad \bar{F} = \frac{F}{S} = \frac{\varepsilon_0 E^2}{4} \bar{r}^2, \quad (5)$$

where $F$ if lift force, N; $p_a$ is average electron pressure, N/m$^2$; $S = \pi r^2$ is cross-section area of tower, m$^2$. The last equation in (5) and many over further equations are more general and suitable for common case. However, we make computation for base tower radius only 10 m. In this case the reader see the real (non relative) data, which allow him to better understand the possibility of electrostatic tower. If the lift force is small, it may be increased by increasing the tower base area.

The computation lift force via altitude for different $E$, $K = 2$ and base $r_0 = 10$ m is presented in fig.4.

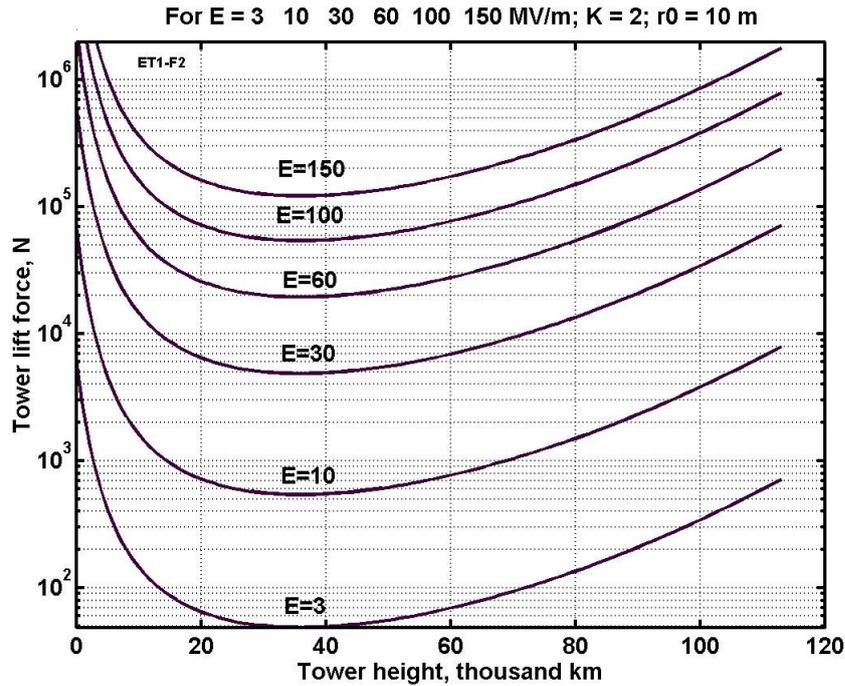

**Fig. 4.** Tower lift force versus tower height for different electric intensity and base radius $r_o = 10$ m and strength coefficient $K = 2$.

As you see for the electric intensity $E = 100$ MV (the dielectric thin film can keep $E = 700$ MV, see Table 2 and below) the electrostatic tower can keep 5 tons if one has altitude at geosynchronous orbit and more 100 tons if one has an altitude 120,000 km.

**4. Dielectric strength of insulator.** As you see above the tower need in film which separate the positive charges located in conductive layer from the electron gas located into tube. This film must have a high dielectric strength. The current material can keep a high $E$ (see table 2 is taken from [10]).

Table 2. Properties of various good insulators (recalculated in metric system)

| Insulator | Resistivity Ohm-m. | Dielectric strength MV/m. $E_i$ | Dielectric constant, $\varepsilon$ |
|---|---|---|---|
| Lexan | $10^{17}$–$10^{19}$ | 320–640 | 3 |
| Kapton H | $10^{19}$–$10^{20}$ | 120–320 | 3 |
| Kel-F | $10^{17}$–$10^{19}$ | 80–240 | 2–3 |
| Mylar | $10^{15}$–$10^{16}$ | 160–640 | 3 |
| Parylene | $10^{17}$–$10^{20}$ | 240–400 | 2–3 |
| Polyethylene | $10^{18}$–$5 \times 10^{18}$ | 40–680* | 2 |
| Poly (tetra- | $10^{15}$–$5 \times 10^{19}$ | 40–280** | 2 |

fluoraethylene)
Air (1 atm, 1 mm gap)                                             4                    1
Vacuum (1.3×10$^{-3}$ Pa,                                      80–120                1
  1 mm gap)

--------------------------------------------------------------------------------------------------

*For room temperature 500 – 700 MV/m.
** 400–500 MV/m.

*Sources:* Encyclopedia of Science & Technology (New York, 2002, Vol. 6, p. 104, p. 229, p. 231) and Kikoin [17] p. 321.

*Note:* Dielectric constant $\varepsilon$ can reach 4.5 - 7.5 for mica (*E* is up 200 MV/m), 6 -10 for glasses (*E* = 40 MV/m), and 900 -3000 for special ceramics (marks are CM-1, T-900) [17], p. 321, (*E* =13 -28 MV/m). Ferroelectrics have $\varepsilon$ up to $10^4$ - $10^5$. Dielectric strength appreciably depends from surface roughness, thickness, purity, temperature and other conditions of materials. Very clean material without admixture (for example, quartz) can have electric strength up 1000 MV/m. As you see we have a needed dielectric material, but it is necessary to find good (and strength) isolative materials and to research conditions which increase the dielectric strength.

**5. Tower cover thickness**. The thickness of tower cover may be found from Equation (3). The result of computation is presented in Fig. 5.

**Fig. 5.** Thickness of tower cover versus tower height for different electric intensity and base radius $r_o$ = 10 m and strength coefficient *K* = 2. (Figure 5 is deleted because size of article is limited 1 Mb)

**6. Mass of tower cover**. The mass of tower cover is

$$dM = 2\pi r \delta \gamma dH, \quad M = \frac{2\pi p r_0^2}{k}\int_0^H \bar{r}^2 dH = \frac{\pi\varepsilon_0 E^2 r_0^2}{k}\int_0^H \bar{r}^2 dH \quad or \quad \bar{M} = \frac{M}{S_0} = \frac{\varepsilon_0 E^2}{k}\int_0^H \bar{r}^2 dH \quad , \quad (6)$$

where *M* is cover mass, kg; $S_0 = \pi r_0^2$ is tower base area, m$^2$, *p* is Eq.(2).

Result of computation is presented in fig. 6.

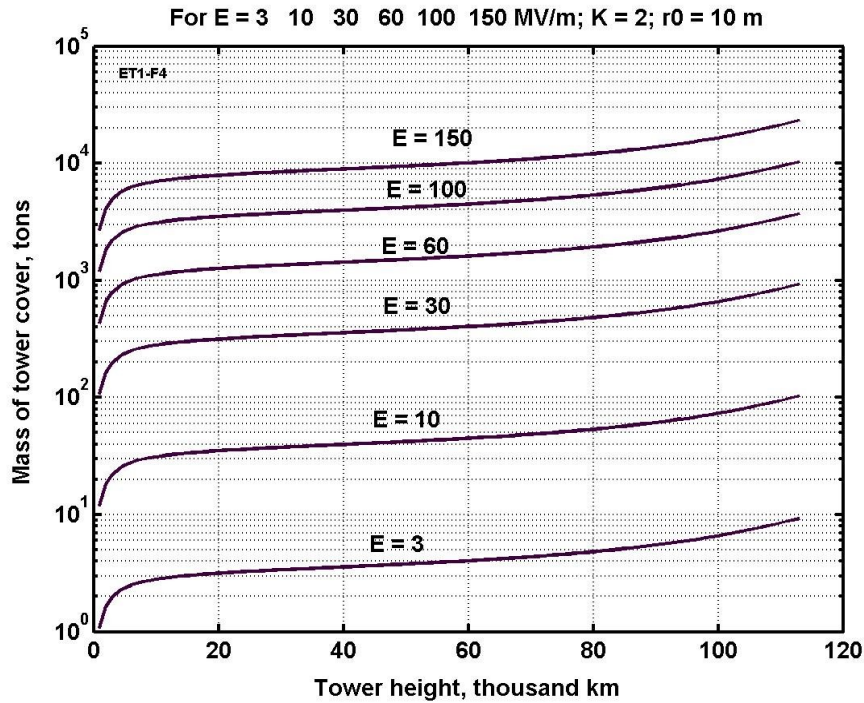

**Fig. 6.** Mass of tower cover versus tower height for a different electric intensity and base radius $r_o$ = 10 m and strong coefficient *K* = 2.

As you see the total mass of 120,000 km electrostatic tower is about 10,000 tons. Compare this number with 3,000,000 tons which has the CNN solid tower in Toronto (Canada) having only 553 m of height.

**7. The volume $V$ and surface of tower $s$** are

$$dV = \pi r^2 dH, \quad V = \pi r_0^2 \int_0^H \bar{r}^2 dH, \quad ds = 2\pi r_0 \bar{r} dH, \quad s = \pi r_0 \int_0^H \bar{r} dH, \qquad (7)$$

where $V$ is tower volume, m$^3$; $s$ is tower surface, m$^2$.

**8. Relation between tower volume charge and tower liner charge** is

$$E_V = \frac{\rho r}{2\varepsilon_0}, \quad E_s = \frac{\tau}{2\pi\varepsilon_0 r}, \quad E_V = E_s, \quad \tau = \pi\rho r^2, \quad \rho = \frac{\tau}{\pi r^2}, \qquad (8)$$

where $\rho$ is tower volume charge, C/m$^3$; $\tau$ is tower linear charge, C/m.

**9. General charge of tower.** We got equation from

$$\tau = 2\pi\varepsilon\varepsilon_0 Er, \quad dQ = \tau dH, \quad Q = 2\pi\varepsilon\varepsilon_0 Er_0 \int_0^H \bar{r} dH, \quad \bar{Q} = \frac{Q}{r_0} = 2\pi\varepsilon\varepsilon_0 \int_0^H \bar{r} dH, \qquad (9)$$

where $Q$ is total tower charge, C; $\varepsilon$ is dielectric constant (see Table 2).

The computation of total charge is shown in fig. 7.

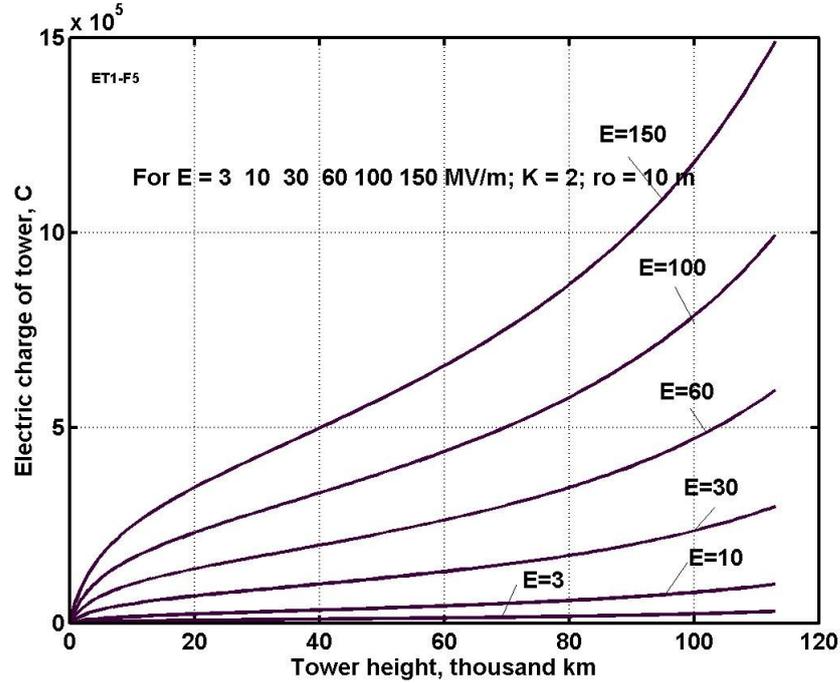

**Fig. 7.** Electric charge of tower versus tower height for different electric intensity and base radius $r_o = 10$ m and strength coefficient $K = 2$.

**10. Charging energy.** The charged energy is computed by equation

$$W = 0.5QU, \quad U = \delta E, \quad W = 0.5Q\delta_a E, \qquad (10)$$

where $W$ is charge energy, J; $U$ is voltage, V. For $E = 100$ MV, $H = 120{,}000$ km, $Q = 12 \times 10^5$ C, $\delta_a = 5 \times 10^{-7}$ m the charged energy is 30 MJ.

**11. Mass of electron gas.** The mass of electron gas is

$$M_e = m_e N = m_e \frac{Q}{e}, \qquad (11)$$

where $M_e$ is mass of electron gas, kg; $m_e = 9.11 \times 10^{-31}$ kg is mass of electron; $N$ is number of electrons, $e = 1.6 \times 10^{-19}$ is the electron charge, C.

The computation for our case give $M_e = 10^{-5}$ kg. That is very small value for gigantic tower-tube 120 thousands km of height.

**12. Power for support of charge.** Leakage current (power) through the cover may be estimated by equation

$$I = \frac{U}{R}, \quad U = \delta E = \frac{r\varepsilon_0 E}{\sigma}, \quad R = \rho\frac{\delta}{s}, \quad I = \frac{sE}{\rho}, \quad W_l = IU = \frac{\delta s E^2}{\rho}, \qquad (12)$$

where *I* is electric currency, A; *U* is voltage, V; *R* is electric resistance, Ohm; $\rho$ is specific resistance, Ohm·m; *s* is tower surface area, m$^2$.

The estimation gives the support power about $0.1 \div 1$ kW.

**13. Electron gas pressure**. The electron gas pressure may be computed by equation (2). This computation is presented in fig. 8.

**Fig. 8.** Electron pressure versus electric intensity (figure 8 is deleted because size of article is limited 1 Mb)

As you see the electron pressure reach 0.5 atm for an electric intensity 150 MV/m and for negligibly small mass of the electron gas.

## Project

As the example (not optimal design!) we take three electrostatic towers having: the base radius $r_0 = 10$ m; $K = 2$; heights $H = 100$ km, 36,000 km (GEO), and $H = 120.000$ km (that may be one tower having named values at given altitudes); electric intensity $E = 100$ MV/m and 150 MV/m. The results of estimation are presented in Table 3.

**Table 3.** The results of estimation main parameters of three AB towers (masts) having the base radius $r_0 = 10$ m and strength coefficient $K = 2$.

| Value | $E$ MV/m | $H$=100 km | $H$=36,000 km | $H$=120,000 km |
|---|---|---|---|---|
| Radius of tower, m | - | 10 | 1 | 4 |
| Useful lift force, ton | 100 | 700 | 5 | 100 |
| Useful lift force, ton | 150 | 1560 | 11 | 180 |
| Cover thickness, mm | 100 | $1 \times 10^{-2}$ | $1 \times 10^{-3}$ | $0.7 \times 10^{-2}$ |
| Cover thickness, mm | 150 | $1.1 \times 10^{-2}$ | $1.2 \times 10^{-3}$ | $1 \times 10^{-2}$ |
| Mass of cover, ton | 100 | 140 | $3 \times 10^3$ | $1 \times 10^4$ |
| Mass of cover, ton | 150 | 315 | $1 \times 10^4$ | $2 \times 10^4$ |
| Electric charge, C | 100 | $1.1 \times 10^4$ | $3 \times 10^5$ | $12 \times 10^5$ |
| Electric charge, C | 150 | $1.65 \times 10^4$ | $4.5 \times 10^5$ | $1.7 \times 10^6$ |

## Conclusion

The offered inflatable electrostatic AB mast has gigantic advantages in comparison with conventional space elevator. Main of them is follows: electrostatic mast can be built any height without rockets, one needs material in tens times less them space elevator. That means the electrostatic mast will be in hundreds times cheaper then conventional space elevator. One can be built on the Earth's surface and their height can be increased as necessary. Their base is very small.

The main innovations in this project are the application of electron gas for filling tube at high altitude and a solution of a stability problem for tall (thin) inflatable mast by control structure.